# GLUON PROPAGATORS AND CONFINEMENT

A.NAKAMURA

*Faculty of Education, Yamagata University*
*Koshirakawa, Yamagata990, Japan*
E-mail: nakamura@kenta.kj.yamagata-u.ac.jp

in collaboration with

H.Aiso, M.Fukuda, T.Iwamiya, T.Nakamura and M.Yoshida

*Computational Sciences Division,*
*National Aerospace Laboratory, chofu-city, Tokyo 182, Japan*

## ABSTRACT

We present SU(3) gluon propagators calculated on $48 \times 48 \times 48 \times N_t$ lattices at $\beta = 6.8$ where $N_t = 64$ (corresponding the confinement phase) and $N_t = 16$ (deconfinement) with the bare gauge parameter,$\alpha$, set to be 0.1. In order to avoid Gribov copies, we employ the stochastic gauge fixing algorithm. Gluon propagators show quite different behavior from those of massless gauge fields: (1) In the confinement phase, $G(t)$ shows massless behavior at small and large $t$, while around $5 < t < 15$ it behaves as massive particle, and (2) effective mass observed in $G(z)$ becomes larger as z increases. (3) In the deconfinement phase, $G(z)$ shows also massive behavior but effective mass is less than in the confinement case. In all cases, slope masses are increasing functions of $t$ or $z$, which can not be understood as addtional physical poles.

## 1. Introduction

The title of this conference, "confinement", is needless to say the most interesting and challenging topics in hadron physics. To clarify the phenomena, many approaches have been tried but to our knowledge no confirmative study of gluon propagators was reported.

There are two reasons why the non-perturbative study of gluon propagators is important to understand the confinement: (1) gluons control the inter-quark dynamics and we expect that their propagators show a peculiar behavior at small momentum region. (2) the confinement of quarks may be understood as a linear force among them but the understanding of the gluon confinement is far from the satisfactory stage. Therefore it gives a new insight for the confinement phenomena if we can measure directly gluon propagators at the confinement and deconfinement phases.

Non-perturbative study of gluon propagators is, however, not easy. Analytical methods have limited power and need some ansatz. Lattice study is also not simple task since (1) it requires the gauge fixing and Gribov copies may distort propagators, and (2) we must measure short and long range regions to study their features and therefore work on a large lattice at high $\beta$.

Here we report our numerical study of gluon propagators in the pure SU(3) gauge theory. Propagators of massless gauge fields must have the form,

$$G_{\mu\nu}(p) \equiv \; <A_\mu(p)A_\nu(-p)>$$
$$= \frac{Z}{p^2}(\delta_{\mu\nu} - (1-\alpha)\frac{p_\mu p_\nu}{p^2}) \tag{1}$$

where $\alpha$ is a gauge parameter. We fourier-transform the $p_0$ sector into $t$,

$$G_{\mu\nu}(p_x, 0, 0; t) \tag{2}$$

The spatial momentum is chosen as

$$\vec{p} = (p_x, 0, 0) = (\frac{2\pi}{N_x}, 0, 0) \tag{3}$$

where $N_x$ is the lattice size along $x$ direction.

We call $G_{xx}$ as a longitudinal propagators and $G_{yy}$ and $G_{zz}$ as transverse ones. The transverse propagators, $G_T$, are independent of the gauge parameter $\alpha$ and behave as having a simple pole, i.e., $\exp(-p_x t)$, if they have the form of Eq. (1). Due to the periodic boundary condition, $G_T$ behaves on the lattice,

$$G_T(t) = C \cosh(p_x(t - N_t/2)) \tag{4}$$

This is our reference form to study propagators, and we shall discuss deviations from it.

## 2. Gribov ambiguity and its study on compact U(1)

In 1972, Gribov pointed out that the gauge fixing in non-Abelian gauge theories is not unique, and argued that the proper choice of a configuration among copies is important for the study of infra-red behavior of the theory.[1] Singer proved rigorously the ambiguity problem for SU(N) on $S^d$, and later Killingback found that the same mathematical argument holds also for SU(N) and U(1) on $T^d$, where $S^d$ and $T^d$ are d-dimensional sphere and torus.[2,3]

Gribov copies were first observed on a lattice in Ref.(4) for the compact U(1), and later on many evidences and analyses for SU(2), SU(3) and U(1) have been reported.[5,6,7,8,9]

The compact U(1) model provides us a good place to test algorithms to measure gauge propagators: We know that at $\beta > 1.0$ they must behave as those of photon, i.e., Eq. (4). And it was found that the copies distort the propagators, i.e., if one simply fixes the gauge with Wilson-Mandula algorithm,[10,11] the propagators do not

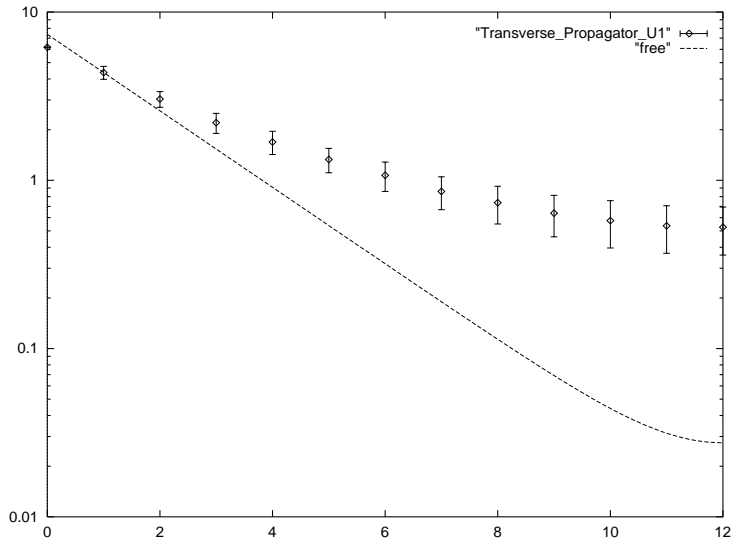

Fig. 1. Transvers "photon" propagator including effects of Gribov copies. $\beta = 1.1$ and the lattice size is $12 \times 8 \times 8 \times 24$. Dotted line is a free massless propagator. Data are taken from Ref.4

behave in a correct manner.[4,9] See Fig.1.

## 3. Stochastic gauge fixing

Nowaday gauge fixing is a standard tool on the lattice, but it is hard to control Gribov copies. One would say that the following definition gives a unique and good gauge condition,[12,15,16]

$$\sum_{x,\mu} Re \text{Tr} U_\mu(x) = Max. \qquad (5)$$

But numerically it is difficult to ensure the above condition is fulfilled especially on large lattices. Several groups have tried to calculate gluon propagators, [12,13,14] but without taking care of the copies.

We have proposed[17,18] to employ the stochastic gauge fixing by Zwanziger, [19,20]

$$\frac{dA_\mu}{d\tau} = -\frac{\delta S}{\delta A_\mu} + \frac{1}{\alpha} D_\mu + \eta_\mu, \qquad (6)$$

where $\tau$ stands for the Langevin time, $\eta$ is a Gaussian noise term. The second term in r.h.s. of Eq. (6) is a force along a gauge trajectory. Usually this force attracts the configuration onto a gauge fixed plane where

$$\Delta \equiv \sum_\mu \partial A_\mu / \partial x_\mu = 0, \qquad (7)$$

and together with the third noise term, the covariant gauge fixed state with $\alpha$ is realized. The algorithm has a remarkable feature, i.e., the second term of Eq.(6) brings the configuration such that

$$\frac{d}{d\tau} \sum_x \Delta^2 \quad < \quad 0 \quad \text{in } \Omega \qquad (8)$$
$$> \quad 0 \text{ out of } \Omega.$$

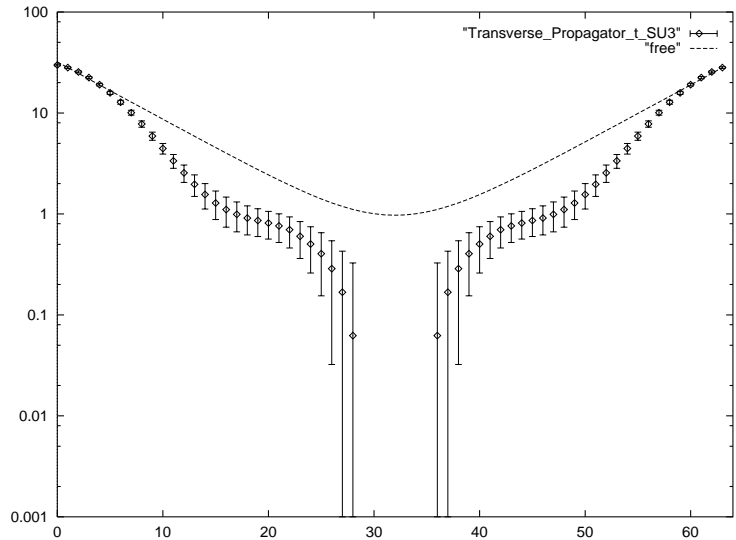

Fig. 2. Transverse gluon propagator, $G_T(t)$, on $48 \times 48 \times 48 \times 64$ at $\beta = 6.8$ (confinement phase)

$\Omega$ is a region where the Faddeev-Popov operator is positive. Namely the stochastic gauge fixing term is attractive (repulsive) inside (outside) of the Gribov region.

A compact lattice version of Eq. (6) was proposed [17] and correct "photon" propagators were observed in the Coulomb phase of compact U(1) model.

### 4. Numerical results

Now we are ready to calculate gluon propagators for SU(3). We use a parallel vector computer, the Numerical Wind Tunnel (NWT) at National Aerospace Laboratory, Japan, which has the peak performance of 236 GFLOPS and 35 GByte memory.

The Langevin time step, $\Delta\tau$, is set to be 0.01. We perform more than 50000 sweeps for each Monte Carlo run, about 10% of which is usually discarded.

In Fig.2 we show the transverse gluon propagator, $G_T(t)$, in the confinement phase, together with the free massless one obtained from Eq.(6). Large deviation appears as $t$ increases; the gluon seems to become massive at large $t$. If we would interpret the behavior as summing several mass poles, some of the coefficients $C$ in Eq. (4), the spectral functions, should be negative. The effective mass seems to vanish again at very large $t$, although there error bars are very large.

We show in Fig.3 and Fig.4 the transverse propagator measured along a spatial direction, $z$, in the confinement and deconfinement phases, respectively. Here effec-tive masses increase monotonically as $z$ increases. In the deconfinement region, this propagator corresponds the screening effect. In the confinement phase, there is no such a screening effect, but the behavior of both propagators is similar. The effective mass in the confinement is larger than in the deconfinement.

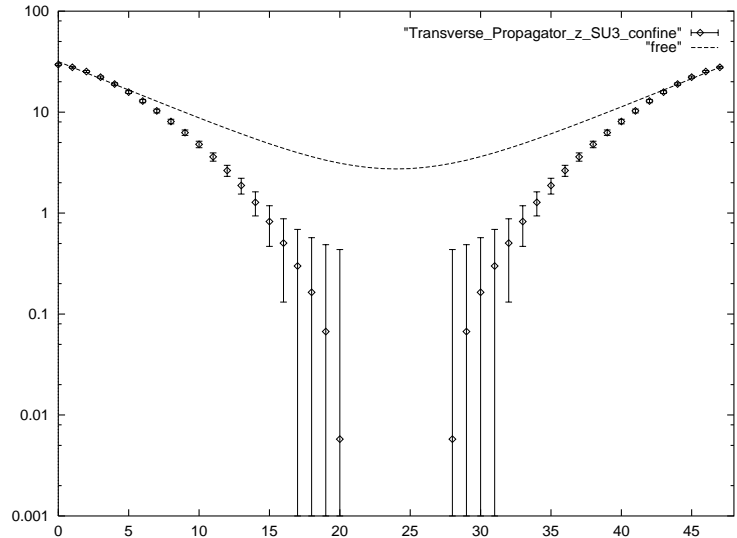

Fig. 3. Spatial transverse gluon propagator, $G_T(z)$, on $48 \times 48 \times 48 \times 64$ at $\beta = 6.8$ (confinement phase)

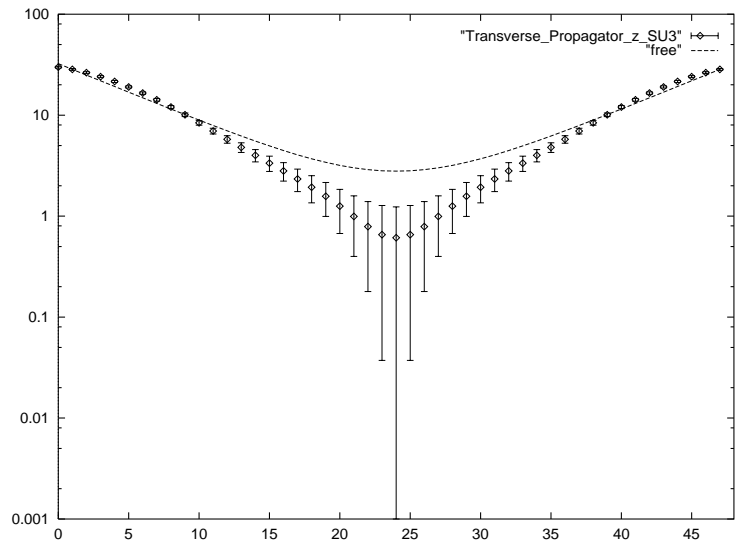

Fig. 4. Spatial transverse gluon propagator, $G_T(z)$, on $48 \times 48 \times 48 \times 16$ at $\beta = 6.8$ (deconfinement phase)

In conclusion we have observed gluon propagators from short to large range. To our knowledge, this is the first measurement of the quantities with taking care of Gribov copies. Data are very encouraging, i.e. error bars are under control and the results are highly non-trivial; we are now collecting high statistical data.

1. V.N.Gribov, *Nucl.Phys.***B139** (1978) 1.
2. I.M.Singer, *Commun.math.Phys.***60** (1978) 7.
3. T.P.Killingback, *Phys.Lett.* **B138** (1984) 87.
4. A.Nakamura and M.Plewnia, *Phys.Lett.***B255** (1991) 274.
5. Ph.deForcrand, J.E.Hetrick, A.Nakamura and M.Plewnia, *Nucl.Phys.***B(Proc.Suppl.)20** (1991) 194-198.
6. E.Marinari, R.Ricci and C.Parrinello, *Nucl.Phys.* **B(Proc.Suppl.)20** (1991) 199-202.
7. A.Hulsebos, M.L.Laursen, J.Smit and A.J.van der Sijs, *Nucl.Phys.* **B(Proc.Suppl.)20** (1991) 199-202.
8. S.Petrarca, *Nucl.Phys.* **B(Proc.Suppl.)26** (1992) 435.
9. V.G.Bornyakov, V.K.Mitrjushikin, M.Muller-Preussker and F.Pahl, Humboldt University Preprint, HU Berlin IEP-93/2.
10. K.G.Wilson, in *Recent Developments in Gauge theories*, ed. G.t'Hooft (Plenum Press, New York, 1980) p.363.
11. J.E.Mandula and M.Ogilvie, *Phys.Lett.*, **B185** (1987) 127.
12. J.E.Mandula and M.Ogilvie, *Phys.Lett.***B185** (1987) 127.
13. C.Bernardm, C.Parrinello and A.Soni, *Phys.Rev.***D49** (1994) 1585.
14. J.Rank, Diplomarbeit, Bielefeld, (1995) March.
15. Semenov-Tyan-Shanskii and Franke, Zapiski Nauchnykh Seminarov, *Leningradskage Otdelleniya Mathemat.* Instituta im Steklov AN SSSR **Vol.120** (1982) 159.
16. T.Maskawa and H.Nakajima, *Prog.Theor.Phys.***63** (1980) 641.
17. A.Nakamura and M.Mizutani, *Vistas in Astronomy* **Vol.37** pp.305-308, 1993, Pergamon Press.
    M.Mizutani and A.Nakamura *Nucl.Phys.* **B(Proc.Suppl.)34** (1994) 253-255.
18. H.Aiso, M.Fukuda, T.Iwamiya, M.Mizutani, A.Nakamura, T.Nakamura and M.Yoshida *Nucl.Phys.* **B(Proc.Suppl.)42** (1995) 899-901.
19. D.Zwanziger, *Nucl.Phys.***B192** (1981) 259.
20. E.Seiler, I.O.Stamatescu and D.Zwanziger, *Nucl.Phys.* **B239** (1984) 177, 201.